\documentclass[aps,prb,twocolumn,showpacs,superscriptaddress,floatfix]{revtex4}

\usepackage{graphicx}
\usepackage{amssymb}

\begin{document}

\title{Thermal expansion of the superconducting ferromagnet UCoGe}

\author{A. Gasparini}
\affiliation{Van der Waals - Zeeman Institute, University of
Amsterdam, Valckenierstraat~65, 1018 XE Amsterdam, The
Netherlands}
\author{Y. K. Huang}
\affiliation{Van der Waals - Zeeman Institute, University of
Amsterdam, Valckenierstraat~65, 1018 XE Amsterdam, The
Netherlands}
\author{J. Hartbaum}
\affiliation{Physikalisches Institut, Karlsruher Institut f{\"u}r
Technologie, Karlsruhe, D-76131, Germany}
\author{H. v. L{\"o}hneysen}
\affiliation{Physikalisches Institut, Karlsruher Institut f{\"u}r
Technologie, Karlsruhe, D-76131, Germany} \affiliation{Institut
f{\"u}r Festk{\"o}rperphysik, Karlsruher Institut f{\"u}r
Technologie, Karlsruhe, D-76021, Germany}
\author{A. de Visser}
\email{a.devisser@uva.nl} \affiliation{Van der Waals -
 Zeeman
Institute, University of Amsterdam, Valckenierstraat~65, 1018 XE
Amsterdam, The Netherlands}

\date{\today}

\begin{abstract}
We report measurements of the coefficient of linear thermal
expansion, $\alpha (T)$, of the superconducting ferromagnet UCoGe.
The data taken on a single-crystalline sample along the
orthorhombic crystal axes reveal a pronounced anisotropy, with the
largest length changes along the $b$ axis. The large values of the
step sizes $\Delta \alpha$ at the magnetic and superconducting
phase transitions provide solid evidence for bulk magnetism and
superconductivity. Specific-heat measurements corroborate bulk
superconductivity. Thermal-expansion measurements in magnetic
fields $B \parallel a,b$ show $\Delta \alpha$ at $T_C$ grows
rapidly, which indicates the character of the ferromagnetic
transition becomes first-order-like.

\end{abstract}


\maketitle

The intermetallic compound UCoGe ($T_s = 0.8$~K and $T_C = 3$~K)
belongs to the small group of superconducting
ferromagnets~\cite{Huy-PRL-2007}. Superconducting ferromagnets
(SCFMs), have the intriguing property that superconductivity (SC)
occurs in the ferromagnetic (FM) phase, at a temperature $T_s$
well below the Curie temperature $T_C$, without expelling magnetic
order~\cite{deVisser-EMSAT-2010}. Until now, this peculiar ground
state has been found in a few materials - all uranium
intermetallics - only: in UGe$_2$~\cite{Saxena-Nature-2000} and
UIr~\cite{Akazawa-JPCM-2004} under pressure, and in
URhGe~\cite{Aoki-Nature-2001} and UCoGe~\cite{Huy-PRL-2007} at
ambient pressure. SCFMs attract much attention, because their
ground state does not obey the standard BCS scenario for
phonon-mediated SC, because the FM exchange field impedes
pairing-up of electrons in spin-singlet Cooper
pairs~\cite{Berk-PRL-1966}. Instead, the itinerant nature of the U
$5f$ magnetic moments, together with the notion that these
materials are close to a magnetic instability, has led to the
proposal that SC is unconventional and promoted by a novel pairing
mechanism~\cite{Fay-PRB-1980,Lonzarich-CUP-1997}: critical FM spin
fluctuations mediate pairing of electrons in spin-triplet states.
In recent years ample evidence for such an unusual pairing
mechanism in SCFMs has been put
forward~\cite{Saxena-Nature-2000,Sandeman-PRL-2003,Levy-NaturePhys-2007,Miyake-JPSJ-2008}.
SCFMs are excellent laboratory tools to investigate the interplay
of magnetism and SC, which is a key issue in unravelling the
properties of a wide range of materials, like heavy-fermion,
high-$T_s$ cuprate and FeAs-based superconductors.

UCoGe crystallizes in the orthorhombic TiNiSi structure (space
group $P_{nma}$)~\cite{Canepa-JALCOM-1996}. The coexistence of SC
and FM in UCoGe was first reported for polycrystalline
samples~\cite{Huy-PRL-2007}. Magnetization measurements show the
emergence of a weak FM phase below $T_C = 3$~K with a small
ordered moment $m_{0}$ = 0.03 $\mu_B$. The analysis of the
magnetization data by means of Arrott plots corroborates itinerant
FM. This is further substantiated by specific-heat data, which
show that the entropy associated with the magnetic phase transition is
small (0.3 \% of $R$ln2). In the FM phase, SC is found with a
transition temperature $T_s = 0.8$~K, as determined by resistance
measurements. The ac-susceptibility shows large diamagnetic
signals below $T_s = 0.6$~K. Thermal-expansion and specific-heat
measurements on polycrystalline samples~\cite{Huy-PRL-2007}
confirmed the bulk nature of the SC and FM phases, with $T_s
^{bulk} = 0.45$~K and $T_C ^{bulk} = 3$~K, respectively.

Since the electronic and magnetic parameters of U$TX$ compounds,
with $T$ a transition metal and $X$ is Si or Ge, are in general
strongly anisotropic~\cite{Sechovsky-handbook-1998} it is of
uttermost importance to carry out further research on high-quality
single-crystalline samples. Recently, Huy {\it et al.} reported
the first magnetic and transport measurements on single
crystals~\cite{Huy-PRL-2008}. Magnetization data revealed FM in
UCoGe is uniaxial, with $m_{0}$= 0.07 $\mu_B$ pointing along the
orthorhombic $c$ axis. Resistance measurements showed the upper
critical field, $B_{c2}(T)$, has an unusual large anisotropy, with
$B_{c2}(T \rightarrow 0)$ for $B
\parallel a,b$ a factor $\sim 10$ larger than for $B
\parallel c$.

In this Brief Report we present measurements of the thermal
properties of UCoGe single crystals. We find that the coefficients
of linear thermal expansion measured along the crystal axes
display a pronounced anisotropy, with the largest length changes
along the $b$ axis. Large values of the step sizes $\Delta \alpha$
at the FM and SC phase transitions provide evidence for bulk
magnetism and SC. Specific-heat measurements support this
conclusion. We use the Ehrenfest relation to analyze the uniaxial
pressure dependencies of $T_s$ and $T_C$. Thermal-expansion
measurements in applied magnetic fields indicate the nature of the
FM phase transition changes to first order.

Single crystals of UCoGe were prepared by the Czochralski method
as described in Refs.~\onlinecite{Huy-PRL-2008,Huy-JMMM-2009}. The
measurements were carried out on two samples, both shaped into a
bar by means of spark erosion, with typical dimensions of $1
\times 1 \times 4$~mm$^3$ and the long direction along the $a$
(sample \#1) and $b$ axis (sample \#2). The samples were
annealed~\cite{Huy-JMMM-2009} and their good quality is attested
by the high residual resistance ratio's,
$RRR=R(300~$K$)/R(1~$K$)$, of $\sim 30$ and $\sim 40$ for sample
\#1 and \#2, respectively. Sample \#1 was previously used to
obtain the data in
Refs~\onlinecite{Huy-PRL-2008,Slooten-PRL-2009}. It shows a large
diamagnetic signal at the superconducting transition, $T_s =
0.5~$K, with a magnitude of 80\% of the ideal screening value. The
coefficient of linear thermal expansion, $\alpha = L^{-1}
(dL/dT)$, was measured using a three-terminal parallel-plate
capacitance method using a sensitive
dilatometer~\cite{deVisser-thesis-1986}. Length changes along the
$a$, $b$ and $c$ crystal axes were measured along the short edges
($\sim 1$~mm) of the samples: $\alpha _b$ and $\alpha _c$ were
measured on sample \#1 and $\alpha _a$ on sample \#2. The data
were taken in a $^3$He system in the $T$-range $0.23-15$~K and in
a dilution refrigerator for $T = 0.05-1$~K. The specific heat was
measured on sample \#2 (mass $\sim 0.1$~g) using a semi-adiabatic
heat-pulse technique for $T = 0.15-1$~K.

\begin{figure}
\includegraphics[width=6.5cm]{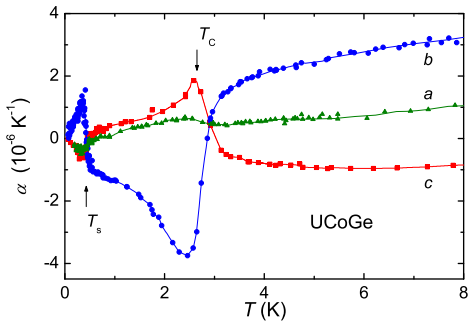} 
\caption{(color online) Coefficient of linear thermal expansion
{\it versus} temperature of UCoGe along the orthorhombic $a$, $b$
and $c$ axis as indicated. Arrows indicate the ferromagnetic (at
$T_C$) and superconducting (at $T_s$) transition temperatures.}
\end{figure}

In Fig.~1 we show $\alpha (T)$ for $T \leq 8$~K measured along the
main crystal axes. The data reveal a strong anisotropy. In the
paramagnetic phase, $\alpha _a$ and $\alpha _b$ are positive, while
$\alpha _c$ is negative. The most pronounced variation is observed
along the $b$ axis. For this direction, the transition to FM yields a negative and
to SC a positive contribution to $\alpha$. For the $a$ and $c$ axis
the contributions are smaller and the polarity is reversed. At the
FM and SC phase transitions large step-like changes, $\Delta
\alpha$, are found. The large values of $\Delta \alpha$ at $T_C$
and $T_s$ provide solid evidence that FM and SC are bulk
properties. Notice, the step sizes at $T_s$ are comparable to the
ones obtained for the heavy-fermion SCs
URu$_2$Si$_2$~\cite{VanDijk-PRB-1995} and
UPt$_3$~\cite{VanDijk-JLTP-1993}. The coefficient of volumetric
expansion is given by $\beta = \sum _i \alpha _i$, where $i = a,
b, c$, and is reported in Fig.~2. Ideally, the $\alpha _i (T)$
curves should be measured on one single sample. However, in our
case we used two samples with slightly different $RRR$ values. The
resulting $\beta (T)$ data shows a large negative step at $T_C$
and a positive step at $T_s$. Since the phase transitions are
relatively broad in temperature, we use an equal area
construction~\cite{equal-volume} to obtain idealized sharp
transitions. In this way we extract $T_C ^{bulk}= 2.6$~K and $T_s
^{bulk}= 0.42$~K. In the inset to Fig.~2 we compare $\beta (T)$ of
the single crystal with previous results on a
polycrystal~\cite{Huy-PRL-2007}, where we assume $\beta = 3 \times
\alpha$. The data show a nice overall agreement, but, obviously,
the phase transitions are much sharper for the single crystal.

\begin{figure}
\includegraphics[width=6.5cm]{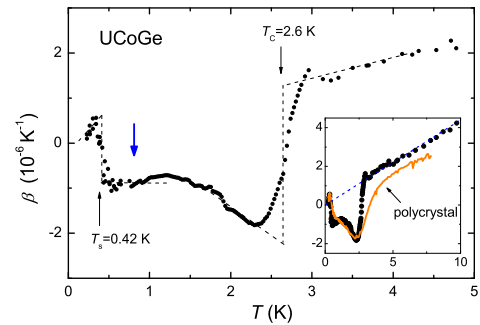}
\caption{(color online) Coefficient of volumetric thermal
expansion of single-crystalline UCoGe as a function of
temperature. The dashed lines represent idealized sharp FM and SC
transitions, at $T_C = 2.6$~K and $T_s = 0.42$~K, respectively.
The blue arrow locates the presence of an additional contribution
in the FM state (see text). Inset: Comparison of $\beta (T)$ of
single-crystalline (closed circles) and polycrystalline (solid
line) UCoGe. The dashed line gives $\beta _{para} (T) = aT$ (see
text). }
\end{figure}

\begin{figure}
\includegraphics[width=5cm]{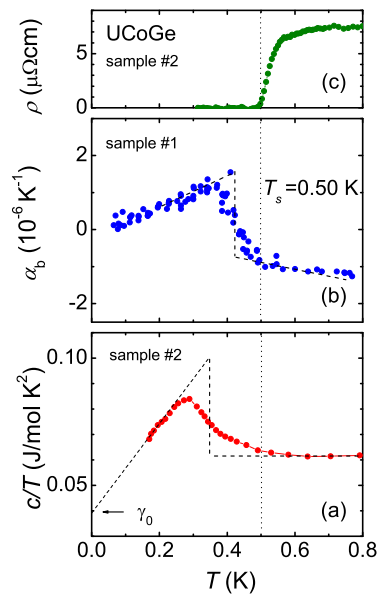}
\caption{(color online) (a) Specific heat of UCoGe (single crystal
\#2) in a plot of $c/T$ {\it versus} $T$. (b) $\alpha _b (T)$ of
UCoGe (single crystal \#1). (c) Resistivity {\it versus} $T$ of
UCoGe (single crystal \#2). The vertical dotted line indicates the
approach to the zero-resistance state coinciding with the onset
temperature of bulk SC as seen in $c/T$ and $\alpha _b (T)$. The
dashed lines in (a) and (b) represent idealized sharp FM and SC
transitions.}
\end{figure}

Specific-heat, $c(T)$, data around the SC transition are reported
in Fig.~3(a). The phase transition for this crystal \#2 is broad,
with $\Delta T_s \sim 0.2$~K. An estimate for the step size
$\Delta (c/T_{s})$ can be deduced using an equal entropy method
(dashed line in Fig.~3a), which yields an idealized transition at
$T_s = 0.35$~K and $\Delta (c/T_{s})/ \gamma _N \approx 0.7$,
where $\gamma _N = 0.062$~J/molK$^2$ is the Sommerfeld
coefficient. This value is considerably smaller than the BCS value
1.43 for a conventional SC. On the other hand, a smooth
extrapolation of $c/T$ versus $T=0$ indicates the presence of
a residual term $\gamma _0 = 0.04$~J/molK$^2$. Since orthorhombic
SCFMs are in principle two-band SCs~\cite{Mineev-PRB-2004}, with
equal spin-pairing triplet states $\vert \uparrow \uparrow
\rangle$ and $\vert\downarrow \downarrow \rangle$ in the spin-up
and spin-down bands, respectively, a finite $\gamma _0$-value
could be taken as evidence that only one band
superconducts~\cite{Belitz-PRB-2004}, in which case $\gamma _0 =
\gamma _N /2$. However, in our case the broad transition and
finite $\gamma _0$-value strongly suggest sample quality is an
issue. The low value $\Delta (c/T_{s})/ \gamma _N$ and finite
$\gamma _0$ term remind one of the early specific-heat data on
single crystals of the heavy-fermion SC
UPt$_3$~\cite{Franse-ZPhys-1985}. Upon improving the sample
quality the transition became more and more sharp, and eventually
a split transition appeared, as well as a much reduced $\gamma _0$
value~\cite{Fisher-PRL-1989}.

In Fig.~3(c) and (b) we compare $c(T)/T$ with resistivity, $\rho
(T)$, data taken on the same sample, and $\alpha _b (T)$ measured
on sample \#1. The resistivity measurements were carried out with
a four-point low-frequency ac-method, with a current of $100~ \mu
$A along the $b$ axis. The zero-resistance state is reached at 0.5
K, which corresponds to the onset temperature $T_s ^{onset}$ for
bulk SC. Using idealized constructions for the SC phase transition
in $c/T$ (Fig.~3a) and $\alpha _b$ (Fig.~3b) we obtain $T_s
^{bulk}$ is 0.35~K and 0.42~K, for sample \#2 and \#1,
respectively.

\begin{figure}
\includegraphics[width=6.5cm]{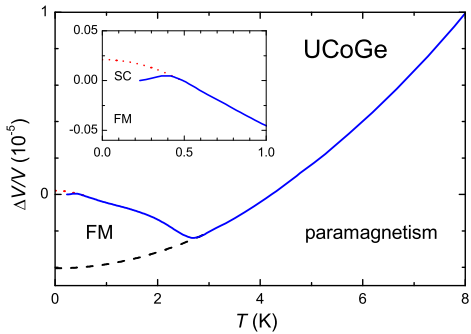}
\caption{(color online) The relative volume change $\Delta V/V =
(V(T)-V(0.05$K$))/V$ as a function of $T$ (solid blue line). The
black dashed line gives $\Delta V/V$ in the absence of FM order.
The red dotted line gives a smooth extrapolation of $\Delta V/V$
in the absence of SC. Inset: Blow-up of the low-$T$ part.}
\end{figure}

\begin{figure}
\includegraphics[width=5cm]{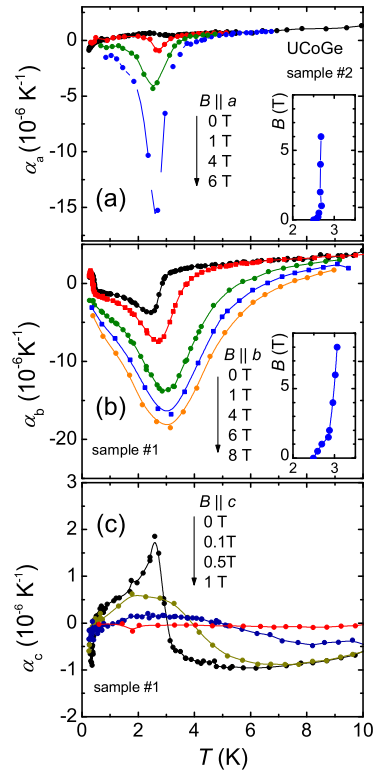}
\caption{(color online) Coefficient of thermal expansion of UCoGe
in applied magnetic fields along the dilatation direction as
indicated. (a) $\alpha _a$ for $B \parallel a$; (b) $\alpha _b$
for $B
\parallel b$; (c) $\alpha _c$ for $B \parallel c$. Insets: $T _C$ as a function of $B \parallel a, b$.}
\end{figure}

With the help of the Ehrenfest relation for second-order phase
transitions $dT_{s,C}/dp_i = V_{m} \Delta \alpha _i / \Delta
(c/T_{s,C})$ (where the molar volume $V_{m} = 3.13 \times 10^{-5}
~$m$^{3}$/mol) one may extract the uniaxial pressure variation of
$T _s$ and $T _C$. Since not all steps $\Delta \alpha _i$ and
$\Delta (c/T_{s,C})$ have been measured on the same sample, we
here restrict ourselves to a qualitative analysis. The largest
effect is calculated for uniaxial pressure, $p _b$, along the $b$
axis: $T _s$ increases and $T _C$ decreases. For $p _a$ and $p _c$
the effect is smaller with reversed polarity. An estimate of the
variation of $T _s$ as a function of hydrostatic pressure can be
calculated using the relation: $dT_{s}/dp = V_{m} \Delta \beta /
\Delta (c/T_{s})$. By combining the results obtained on the two
crystals, using the values $\Delta \beta = 1.19 \times
10^{-6}$~K$^{-1}$ (see Fig.~2) and $\Delta c/T _s=
0.038$~J/molK$^2$ (Fig.~3), we calculate $dT_{s}/dp = 0.098$
K/kbar. This value is larger than the value $0.062$ K/kbar deduced
for a polycrystal~\cite{Huy-PRL-2007,remark_Ehrenfest}. In the
same way we calculate $dT_{C}/dp = -0.79$ K/kbar, where we used
$\Delta \beta = -3.53 \times 10^{-6}$~K$^{-1}$ (Fig.~2) and the
polycrystal value $\Delta c/T _C = 0.014$~J/molK$^2$ (Ref.
\onlinecite{Huy-PRL-2007}). Notice, the pressure variations
deduced from the Ehrenfest relation are considerably larger than
the experimental values $dT_{s}/dp = 0.03$~K/kbar and $dT_{C}/dp
\backsim
-0.21$~K/kbar~\cite{Slooten-PRL-2009,Hassinger-JPSJ-2008}, which
tells us the quantitative analysis should be interpreted with
care.

The relative volume changes due to FM order and SC are obtained by
integrating $\beta (T)$ versus $T$. The result is shown in Fig.~4.
The spontaneous magnetostriction is obtained by integrating
$\beta_{FM} (T)$, {\it i.e.}, the difference between the measured
$\beta (T)$ and the paramagnetic background term. The latter is
approximated by a linear term $\beta _{para} = aT$ with $a = 4.4
\times 10^{-7}$ K$^{-2}$ (see inset Fig.~2). The relative volume
change due to the spontaneous magnetostriction amounts to $\Delta
V/V = 4.2 \times 10^{-6}$ for $T \rightarrow 0$ and is much larger
(and has an opposite sign) than the estimated $\Delta V /V = - 2.5
\times 10^{-7}$ due to SC (see inset Fig.~4). The latter value is
due to the condensation energy of the SC state and agrees well
with similar values obtained for heavy-fermion
superconductors~\cite{VanDijk-PRB-1995,VanDijk-JLTP-1993}. Thus FM
order is not expelled below $T_s$ and coexists with
superconductivity. $\mu$SR experiments~\cite{DeVisser-PRL-2009}
provide evidence for the coexistence of SC and FM on the
microscopic scale.

A closer inspection of the volumetric thermal expansion in Fig.~2
reveals an additional contribution visible below $\sim 1.5$~K in
the FM phase, just before SC sets in. This shoulder indicates the
presence of a second energy scale, most likely related to
low-energy spin fluctuations. It will be highly interesting to
investigate whether these spin fluctuations provide the pairing
interaction for SC. Notice, a second low-energy scale associated
with spin fluctuations has also been identified in the thermal
expansion and specific heat of URhGe and
UGe$_2$~\cite{Sakarya-PRB-2003,Hardy-PRB-2009}.

Finally, we present measurements of $\alpha (T)$ around the Curie
point in magnetic fields applied along the dilatation direction
(see Fig.~5). Again we observe a large anisotropy. For $\alpha _c$
and $B \parallel c \parallel m_0$ the phase transition smears out
rapidly: in a field of 1 T, $\alpha _c (T)$ is virtually
independent of temperature up to 10~K and close to zero. For $B
\parallel a,b$ the magnetic contribution to $\alpha _a$ and
$\alpha _b$ grows rapidly, and attains the large values of $\sim
-2 \times 10 ^{-5}$~K$^{-1}$ at $T _C$ in a field of 8~T. The
large length changes show the nature of the FM transition becomes
first-order-like in an applied magnetic field. This is in line
with the phase diagram for an itinerant quantum FM when tuned to
the critical point~\cite{Belitz-PRL-2005} with the magnetic field
playing the role of pressure. A recent analysis of the Landau free
energy of FM UCoGe in a magnetic field predicts $T_C$ is reduced
in a transverse field $B \perp m_0$~\cite{Mineev-arXiv-2010}. The
variation of $T_C$ with magnetic field $B \parallel a, b$ as
determined from the thermal expansion data in field is given in
the insets of Figs~5a and b, respectively. $T_C$ shows a small
increase in low magnetic fields, but then is rather insensitive
for $B$ up to 8 T. Magnetotransport data reveal that for $B
\parallel b$ the critical field at which
$T_C \rightarrow 0$ is $\sim$15 T~\cite{Aoki-JPSJ-2009}.

In summary, we have investigated the thermal properties of the
SCFM UCoGe. The use of single-crystalline samples enabled us to
investigate the anisotropy in the coefficient of the linear
thermal expansion. The largest length changes, $\Delta L/L$ are
observed along the $b$ axis. Large phase-transition anomalies at
$T _C$ and $T _s$ confirm bulk magnetism and bulk SC. By making
use of Ehrenfest relations, the effect on $T _C$ and $T _s$ of
uniaxial pressure was investigated. In the volumetric thermal
expansion an additional contribution was observed which develops
toward low $T$, just before SC sets in. Experiments on large,
high-quality single crystals are required to further investigate
this phenomenon as it may provide an important clue as regards
low-energy spin fluctuations providing the glue for
superconductivity.

This work was part of the research programme of the Foundation for
Fundamental Research on Matter (FOM), which is financially
supported by the Netherlands Organisation for Scientific Research
(NWO), and of the EC 6th Framework Programme COST Action P16 ECOM.
Support by the Deutsche Forschungsgemeinschaft under FOR 960 is also acknowledged.


\begin{thebibliography}{99}

\bibitem{Huy-PRL-2007} N. T. Huy, A. Gasparini, D. E. de Nijs, Y. Huang, J. C. P. Klaasse,
T. Gortenmulder, A. de Visser, A. Hamann, T. G\"{o}rlach, and H.
von L\"{o}hneyesen, Phys. Rev. Lett. {\bf 99}, 067006 (2007).

\bibitem{deVisser-EMSAT-2010} A. de Visser, {\it Superconducting
Ferromagnets}, in Encyclopedia of Materials: Science and
Technology, eds K. H. J. Buschow {\it et al.} (Elsevier, Oxford,
2010), pp. 1-6.


\bibitem{Saxena-Nature-2000} S. S. Saxena, P. Agarwal, K. Ahilan,
F. M. Grosche, R. K. W. Haselwimmer, M. J. Steiner, E. Pugh, I. R.
Walker, S. R. Julian, P. Monthoux, G. G. Lonzarich, A. Huxley, I.
Sheikin, D. Braithwaite, and J. Flouquet, Nature (London) {\bf
406}, 587 (2000).

\bibitem{Akazawa-JPCM-2004} T. Akazawa,
H. Hidaka, T. Fujiwara, T. C. Kobayashi, E. Yamamoto, Y. Haga, R.
Settai, and Y. {\={O}}nuki,
 J. Phys.: Condens. Matter {\bf 16}, L29 (2004).

\bibitem{Aoki-Nature-2001} D. Aoki, A. Huxley, E. Ressouche,
D. Braithwaite, J. Flouquet, J. P. Brison, E. Lhotel, and C.
 Paulsen,
Nature (London) {\bf 413}, 613 (2001).


\bibitem{Berk-PRL-1966} N. F. Berk, and J. R. Schrieffer,
Phys. Rev. Lett. {\bf 17}, 433 (1966).

\bibitem{Fay-PRB-1980} D. Fay, and J. Appel,
Phys. Rev. B {\bf 22}, 3173 (1980).

\bibitem{Lonzarich-CUP-1997}
G. G. Lonzarich, in {\it Electron: A Centenary Volume}, ed. M.
Springford (Cambridge Univ. Press, Cambridge, 1997), Chapter 6.

\bibitem{Sandeman-PRL-2003} K. G. Sandeman, G. G. Lonzarich, and A. J. Schofield,
Phys. Rev. Lett. {\bf 90}, 167005 (2003).

\bibitem{Levy-NaturePhys-2007} F. L\'{e}vy, I. Sheikin, and A. Huxley,
Nature Physics {\bf 3}, 460 (2007).

\bibitem{Miyake-JPSJ-2008} A. Miyake, D. Aoki, and J. Flouquet,
J. Phys. Soc. Jpn {\bf 77}, 094709 (2008).

\bibitem{Canepa-JALCOM-1996} F. Canepa, P. Manfrinetti, M. Pani, and A. Palenzona,
J. Alloys Comp. {\bf 234}, 225 (1996).

\bibitem{Sechovsky-handbook-1998} V. Sechovsk{\'{y}}, and L.
Havela, {\it Handbook of Magnetic Materials} Vol.~11 ed. K. H. J.
Buschow (North Holland, Amsterdam, 1998) pp.~1-289.

\bibitem{Huy-PRL-2008} N. T. Huy, D. E. de Nijs, Y. K. Huang, and A. de Visser,
Phys. Rev. Lett. {\bf 100}, 077002 (2008).

\bibitem{Huy-JMMM-2009} N. T. Huy, Y. K. Huang, and A. de Visser, J. Magn. Magn. Mat. {\bf 321},
2691 (2009).

\bibitem{Slooten-PRL-2009} E. Slooten, T. Naka, A. Gasparini, Y. K. Huang, and A. de
Visser,Phys. Rev. Lett. {\bf 103}, 097003 (2009).

\bibitem{deVisser-thesis-1986} A. de Visser,
{\it Ph.D. Thesis} (University of Amsterdam, 1986), unpublished.

\bibitem{VanDijk-PRB-1995} N. H. van Dijk, A. de Visser, J. J. M. Franse, and A. A. Menovsky,
Phys. Rev. B {\bf 51}, 12665 (1995).

\bibitem{VanDijk-JLTP-1993} N. H. van Dijk, A. de Visser, J. J. M. Franse, and L. Taillefer,
J. Low Temp. Phys. {\bf 93}, 101 (1993).

\bibitem{equal-volume} An equal volume for the broadened and
idealized contributions is imposed when integrating $\beta (T)$
with respect to the background signal.

\bibitem{Mineev-PRB-2004} V. P. Mineev, and T. Champel,
Phys. Rev. B {\bf 69}, 144521 (2004).

\bibitem{Belitz-PRB-2004} D. Belitz, and T. R. Kirkpatrick,
Phys. Rev. B {\bf 69}, 184502 (2004).

\bibitem{Franse-ZPhys-1985} J. J. M. Franse, A. Menovsky, A. de Visser, C. D. Bredl, U.
Gottwick, W. Lieke, H. M. Mayer, U. Rauchschwalbe, G. Sparn, and
F. Steglich, Z. Phys. B {\bf 59}, 15 (1985).

\bibitem{Fisher-PRL-1989} R. A. Fisher, S. Kim, B. F. Woodfield, N. E. Phillips, L. Taillefer,
 K. Hasselbach, J. Flouquet, A. L. Giorgi, and J. L. Smith,
Phys. Rev. Lett. {\bf 62}, 1411 (1989).

\bibitem{remark_Ehrenfest} The value of $dT_{s}/dp$ in Ref.1 should read
$0.020$~K/kbar. Notice, this value and $dT_{C}/dp = -0.25$~K/kbar
refer to uniaxial pressure dependencies, since they are evaluated
using $\Delta \alpha$ rather than $\Delta \beta$.

\bibitem{Hassinger-JPSJ-2008} E. Hassinger, D. Aoki, G. Knebel, and J. Flouquet,
J. Phys. Soc. Jpn {\bf 77}, 073703 (2008).

\bibitem{DeVisser-PRL-2009} A. de Visser, N. T. Huy,
A. Gasparini, D. E. de Nijs, D. Andreica, C. Baines, and A. Amato,
Phys. Rev. Lett. {\bf 102}, 167003 (2009).


\bibitem{Sakarya-PRB-2003} S. Sakarya, N. H. van Dijk, A. de Visser, and E.
Br\"{u}ck, Phys. Rev. B {\bf 67}, 144407 (2003).

\bibitem{Hardy-PRB-2009} F. Hardy, C. Meingast, V. Taufour, J.
Flouquet, H. v. L\"{o}hneysen, R. A. Fisher, N. E. Phillips, A.
Huxley, and J. C. Lashley, Phys. Rev. B {\bf 80}, 174521 (2009).

\bibitem{Belitz-PRL-2005} D. Belitz, T. R. Kirkpatrick, and J.
Rollb\"{u}hler, Phys. Rev. Lett. {\bf 94}, 247205 (2005).

\bibitem{Mineev-arXiv-2010} V. P. Mineev, e-print arXiv:1002.3510v1.

\bibitem{Aoki-JPSJ-2009} D. Aoki, T. D. Matsuda, V. Taufour, E. Hassinger,
G. Knebel, and J. Flouquet, J. Phys. Soc. Jpn {\bf 78}, 113709
(2009).









\end{thebibliography}
\end{document}